\documentclass{aa}
\usepackage{graphics}
\input{psfig}
\begin{document}

\thesaurus{06(06.05.1, 06.06.4, 06.09.1)}

\headnote{Research Note}

\title{About the time of evolution of a Solar Model}

\author{P. Morel, J. Provost \and G. Berthomieu}

\institute{D\'epartement Cassini, UMR CNRS 6529, Observatoire de la C\^ote 
d'Azur, BP 4229, 06304 Nice CEDEX 4, France}

\offprints{P. Morel}
\mail{morel@obs-nice.fr}

\date{Received date / Accepted date}


\maketitle

\begin{abstract}
The evolution of a solar model is initialized with
homogeneous models of either,
pre-main sequence (P-models) or, zero-age main sequence (Z-models).
The zero-age of a solar model is conventionally referenced as
the time where the nuclear reactions
just begin to dominate gravitation as the primary source of energy.
Fixing the physics, we found that the structure of P- and Z-models computed
with the same physics
are almost similar soon after the exhaustion
of their convective core. This similarity gives  a
connection between the age of the Sun
$t_\odot$ and the time $t_{\rm cal}$ elapsed in the
calculation of calibrated solar models. We found that
a Z-model 
calibrated with  $t_{\rm cal}=t_\odot$ and
a P-model  calibrated 
with $t_{\rm cal}=t_\odot+25$\,My, are indistinguishable at the
relative accuracy level of a few $10^{-4}$.

\end{abstract}

\section{Introduction}\label{sec:int}
The evolution sequences of a solar model are initialized, either with
a chemically homogeneous pre-main sequence model (P-models) powered only
by gravothermal energy, or with a fictitious homogeneous zero-age
main sequence model (Z-models) powered only by thermonuclear energy.
Fixing the physics, whatever they start from, calibrated
solar P- or Z-models have, by definition, the same radius, luminosity and surface 
mass fraction of heavy elements.
If one is not concerned with specific pre-main sequence
physical processes, e.g. primeval mass and angular momentum losses or
lithium depletion, the amount of calculations is half as large
for Z-models (see e.g. Morel  et al. \cite{mpb97}).
The initial internal structure of initial P- and Z-models differs,
despite the same surface parameters at solar age.
It is not obvious that the internal structure of the resulting calibrated models
is alike and, if so, is it just for the resulting models
or also for parts of their previous evolution?
It is important also to establish
a relationship between the age of the Sun and the sum of the time steps
elapsed in the calculation.
As an example, to take the time of the pre-main sequence evolution into account,
Weiss \& Schlattl (\cite{ws98}) have evolved their calibrated solar P-models
 40\,My beyond the age they have assigned to the Sun.

A preliminary discussion is given in Morel et al. (\cite{mpb98}).
The paper is organized as follows: in Sec.~\ref{sec:age} we recall
how the age of a solar model is
defined with respect to the age of the Sun.
Section~\ref{sec:ana} is devoted to our analysis. Results are given in
Sec.~\ref{sec:res} and we conclude in Sec.~\ref{sec:dis}.

\section{The solar age}\label{sec:age}
In a sequence of evolving P-models,
the zero-age main sequence (ZAMS) model is usually defined as
 the first model where $\epsilon_{\rm nuc}$, the nuclear energy generation 
{\em just begin} to dominate $\epsilon_{\rm g}$, the gravothermal energy
liberation (Guenther \& Demarque \cite{gd97}). 
 That definition influences the various age definitions.
Here, with this definition, for a sequence of solar P-models
the age of the {\em calibrated} P-model will be defined as
the evolution time i.e., the sum of the time steps,
from the ZAMS model to the present day model. As we shall
see that definition influences slightly the various age definitions.

Conventionally, the solar age $t_\odot$ is the time it has taken the Sun to
evolve from ZAMS to present day (Guenther \& Demarque \cite{gd97}).
According to a reasonable hypothesis, $t_\odot$ needs to be consistent with
the age of the oldest meteorites $4566\pm5$\,My
(Bahcall et al. \cite{bpw95}). Guenther \& Demarque (\cite{gd97}) have argued
that the radioactive
clocks of these meteorites are zeroing during the last
high-temperature event in the primordial solar system nebula.
This has occurred,
according to Guenther (\cite{g89})$, 40\pm 10$\,My before the ZAMS, therefore
Guenther \& Demarque ($loc.\,cit.$) estimate the ``meteoritic solar age''
to be $t_{\odot\rm m}=4530\pm40$\,My. 
The determination of $t_\odot$ has been recently revisited using
new methods of investigation including statistical arguments
based on helioseismological data.
In their pioneering work Weiss \& Schlattl (\cite{ws98}) investigated how the
known deficits of the solar models would translate into an age, if the
best-fitting model's age  is assumed to be the solar age. They found
a ``helioseismic solar age'' in the interval
$4650\,{\rm My}\la t_{\odot\rm h} \la 5650\,{\rm My}$.
Recently, another analysis of Dziembowski et al. (\cite{dfrs99}),
based only on central helioseismic data,
leads to $t_{\odot\rm h}=4660\pm110$\,My, a value marginally consistent with
$t_{\odot\rm m}$.

\section{Solar P- and Z-models}\label{sec:ana}
Hereafter we shall note $t_{\rm ev}$ the evolution time of any model.
For a calibrated solar model we shall note
$t_{\rm cal}= t_{\rm ev}$. We shall extend
these notations to the Sun itself by setting $t_{\rm ev}=0$ at the time of
the last high-temperature event in the primordial solar system nebula, thus
$t_{\rm cal}=t_{\odot\rm m}+40$\,My (Guenter $loc.\,cit.$).

For a sequence of P-models we conventionally set $t_{\rm ev}=0$ at
the ignition of deuterium; that occurs as soon as
the temperature reaches $T\approx 0.5$\,MK. In such physical conditions 
the model is fully convective and then homogeneous.
At $t_{\rm ev}\approx0.3$\,My, all the initial \element[][2]{H}
is already converted into \element[][3]{He}
via the reaction $\element[][2]{H}(p,\gamma)\element[][3]{He}$ and
$0<\epsilon_{\rm nuc}<<\epsilon_{\rm g}$.   
Therefore, long before the ZAMS, the abundance of \element[][3]{He} is 
enhanced by a factor of $\approx4$ (in mass) with respect to its primeval value,
namely from $\approx2.3\,10^{-5}$ to $\approx8.6\,10^{-5}$.
At $t_{\rm ev}\approx1.2$\,My, the central temperature approaches
$T_{\rm cal}\sim3$\,MK, and the PP burning of
\element[][1]{H} begins in the core. Soon after,
due to the sudden energy liberation, the temperature increases,
the opacity decreases and 
the core becomes radiative. At $t_{\rm ev}\approx 13$\,My
the innermost limit of the solar convection zone has receded to
about its present day location $R_{\rm ZC}\sim0.7R_\odot$;
the central temperature has jumped to $T_{\rm cal}\approx13$\,MK.
The zero-age main-sequence occurs at $t_{\rm ev}=t_{\rm ZAMS}\approx 25$\,My
just when $\epsilon_{\rm nuc}\gse\epsilon_{\rm g}$.
At $t_{\rm ev}\approx 37$\,My an extra energy generation due to the CNO burning 
of \element[][12]{C} into \element[][14]{N} 
creates a convective core. It lasts until
$t_{\rm ev}\approx90$\,My
when the CNO reactions reach their equilibrium state.
Beyond $t_{\rm ev}\goa 50$\,My more than 99\% of the energy
generated have a nuclear origin. 
After the exhaustion of the convective core, the model evolves quietly without
fundamental modifications of its structure until present day.

Therefore, for a calibrated solar P-model
$\tau_{\rm p}= t_{\rm cal}$ writes:
\begin{equation}\label{eq:tp}
\tau_{\rm p}=t_{\rm ZAMS}+t_\odot.
\end{equation}

For a sequence of Z-models we conventionally assign $t_{\rm ev}=0$\,My
to the first model.
In such a model the energy only comes from
nuclear reactions i.e., $\epsilon_{\rm g}=0$. It is an  
idealized model,  because such a homogeneous state cannot be issued from any
previous PMS evolution. As the time goes on, the nuclear burning
and the gravity relax the initial inconsistencies of chemical composition,
temperature, density and pressure.
At $t_{\rm ev}=0$ a solar Z-model presents a convective core
which lasts until $t_{\rm ev}\approx 65$\,My,
when the CNO bi-cycle begins to work at equilibrium. Then, as for the P-models,
after the exhaustion of the convective core, the model changes quietly
until present day.

\begin{figure*}
\centerline{
\psfig{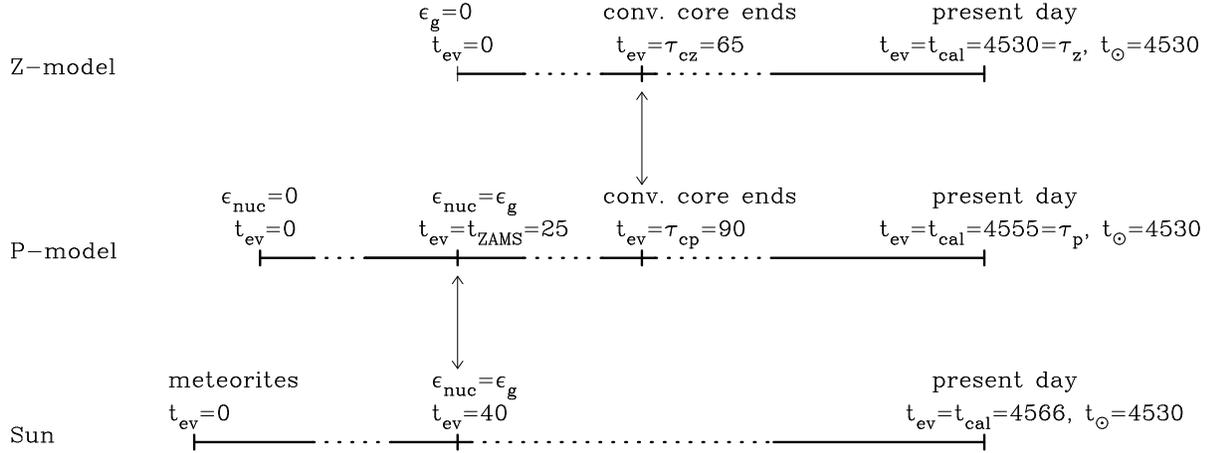}
}
\caption{
Schematic illustrations of chronologies with $t_\odot=t_{\odot \rm m}$
for the Sun, P- and Z-models; the times are in My.
}\label{fig:crobar}
\end{figure*}

The definition of the age
given in Sec.~\ref{sec:age} does not apply to Z-models, since
the gravothermal energy generation {\em never}
overcomes the nuclear one. Thus, the time of the ZAMS
is undefined for a sequence of Z-models. To connect $t_\odot$ and 
$\tau_{\rm z}= t_{\rm ev}$,
{\em we infer} that a P- and a Z-model computed with the same physics
are almost identical
as soon as the nuclear reactions start to work at equilibrium. That occurs
$\approx 10$\,My after the exhaustion of the convective cores. Beyond this
epoch the models will remain alike. As illustrated in
Fig.~\ref{fig:crobar}, calibrated P- and Z-models will have the same
age if the same amount of time is elapsed between present day and 
the epochs $\tau_{\rm cp}$ and $\tau_{\rm cz}$ of
convective core exhaustion:
$\tau_{\rm p}-\tau_{\rm cp}=\tau_{\rm z}-\tau_{\rm cz}$.
Owing to Eq.~(\ref{eq:tp}), $\tau_{\rm z}$ writes:
\begin{equation}\label{eq:t}
\tau_{\rm z}=\tau_{\rm p}-\tau_{\rm cp}+\tau_{\rm cz}=
t_{\rm ZAMS}+t_\odot-\tau_{\rm cp}+\tau_{\rm cz}.
\end{equation}

\section{Results}\label{sec:res}
Basically, the physics used in the calculations is the same as in
Morel et al. (\cite{mpb97}), it uses OPAL opacities and equation of state.
The microscopic diffusion of chemicals is allowed for.
The models are calibrated, within
a relative accuracy better than $10^{-4}$, by adjusting:
the ratio $l/H_{\rm p}$ of the mixing-length to the 
pressure scale height, the initial mass fraction $Y_{\rm ini}$
of helium and the initial mass fraction $(Z/X)_{\rm ini}$ of
heavy elements to hydrogen, so that, at present day,
the models have the luminosity $L_\odot=3.846\,10^{33}$\,erg\,s$^{-1}$,
the radius $R_\odot=6.9599\,10^{10}$\,cm, 
and the mass fraction of heavy element to hydrogen $(Z/X)_\odot=0.0245$.
We have taken into account the 
most important nuclear reactions of PP+CNO cycles
with the species $^2$H, $^7$Li, $^7$Be at equilibrium.
As \element[][2]{H} is set at equilibrium,
the protosolar \element[][2]{H}
is included into the initial \element[][3]{He}; one has for
the initial isotopic ratio 
\element[][3]{He}/\element[][4]{He} $=4.19\,10^{-4}$ (in number)
(Gautier \& Morel \cite{gm97}).
The recently updated rates of Adelberger et al. (\cite{al98}) are used.
The models have been computed using the CESAM code (Morel \cite{m97}).
The evolution of a Z-model necessitates about 40 time
steps whereas 90 for a P-model. Around epochs of
convective core exhaustion and ZAMS, the time step is refined in
order to define these instants with an accuracy better than $\pm2$\,My.

\begin{figure}
\centerline{
\psfig{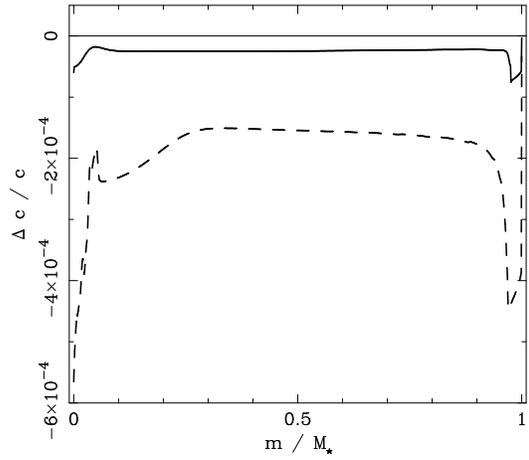}
}
\caption{
Relative differences in sound speed between the P and Z-models Sp and Sz;
dashed: 10\,My after the convective core exhaustion, full: present day. 
}\label{fig:dvson}
\end{figure}

We have adjusted $\tau_{\rm p}$ for the calibrated
P-model Sp so that its age was $t_\odot=4530$\,My.
The ZAMS occurred at $t_{\rm ZAMS}=25\pm2$\,My and the
convective core was exhausted at $\tau_{\rm cp}=90\pm2$\,My.
With the calibration parameters {\em obtained} for Sp,
we have computed the evolution of the Z-model Sz.
We have found that the exhaustion of its convective core occurred at
$\tau_{\rm cz}=65\pm2$\,My.
According to our analysis, after
$\tau_{\rm z}=t_\odot\pm6$\,My, given by Eq.~(\ref{eq:t}),
the Z-model Sz is expected to be
{\em calibrated} and to have a structure close
to the P-model Sp. That is indeed the case, as seen in 
Table~\ref{tab:1} and Fig.~\ref{fig:dvson},
as far as the global
parameters and sound speed profiles are concerned.

\begin{table}
\caption[]{
Comparison of global characteristics of solar P-models (labels p and P)
and Z-models (labels z and Z). The subscripts
``$\rm _s$'' and ``$\rm _c$'' respectively designate a surface and a center
value. The data are for
the time $t_{\rm ev}=0$ (first group), for the time 10\,My
after the epoch $t_{\rm cv}$
of convective core exhaustion (second group) and for $t_{\rm cal}$ i.e.
present day 
(last group). The units are My for time, $10^7$K for the temperatures
and g\,cm$^{-3}$ for the densities.
}\label{tab:1}
\centerline{
\begin{tabular}{llllllllll} \\  \hline \\
                           & Sp          &Sz          &SP          &SZ  \\
\\ \hline \\
$Y_{\rm s}$                &0.2731       &0.2731      &0.2720      &0.2720 \\
$(Z/X)_{\rm s}$            &0.0278       &0.0278      &0.02775     &0.02775 \\
\\ \hline \\
$t_{\rm cv}$		   &$90\pm2$   &$65\pm2$  &$90\pm2$  &$64\pm2$ \\           
$Y_{\rm s}$                &0.2724 &0.2725 &\\
$Z_{\rm s}$                &0.0196 &0.0196 &\\
$R_{\rm ZC}$               &0.7220 &0.7221 &\\
 $T_{\rm c}$               &1.347  &1.349  &\\
$\rho_{\rm c}$             &81.35  &81.31  &\\
$Y_{\rm c}$                &0.2781 &0.2778 &\\
$Z_{\rm c}$                &0.0202 &0.0202 &\\
\\ \hline \\
$t_{\rm ev}$               &4555         &4530        &4685        &4660   \\
$t_\odot$		   &4530         &4530        &4660        &4660     \\
$1-R_{\rm s}/R_\odot$	           &$-5\,10^{-5}$&$6\,10^{-5}$&$6\,10^{-5}$&$-7\,10^{-7}$\\
$1-L_{\rm s}/L_\odot$	           &$-2\,10^{-5}$&$4\,10^{-5}$&$5\,10^{-5}$&$-9\,10^{-5}$\\
$Y_{\rm s}$                &0.2446       &0.2446      &0.2433      &0.2433      \\
$(Z/X)_{\rm s}$            &0.0245       &0.0245      &0.0245      &0.0245     \\
$R_{\rm ZC}/R_\odot$       &0.7137       &0.7137      &0.7121      &0.7122     \\
 $T_{\rm c}$               &1.569        &1.569       &1.572       &1.572      \\
$\rho_{\rm c}$             &152.2        &152.2       &154.4       &154.3       \\
$Y_{\rm c}$                &0.6396       &0.6395      &0.6467      &0.6465      \\
$Z_{\rm c}$                &0.0210       &0.0210      &0.0210      &0.0210      \\
\\ \hline \\
\end{tabular}
}
\end{table}

\begin{figure}
\resizebox{7cm}{!}{\includegraphics{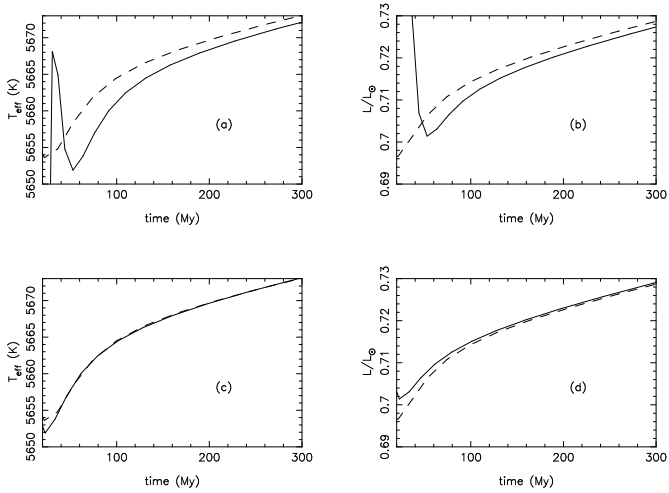}}
\caption{
(a) Effective temperature and (b) luminosity
with respect to time, for the solar P- and Z-models, Sp (full) and Sz (dashed);
(c) and (d) the same with a +25\,My abscissa shift for Sz.
}\label{fig:shift}
\end{figure}

Figure~\ref{fig:shift} shows diagrams of effective temperature
and luminosity versus time. As expected,
an abscissa shift of $+25$\,My for the Z-model
Sz allows to superimpose the loci of Sz and Sp beyond
$t_{\rm ev}\approx 100$\,My. 
We also found that the value of $\tau_{\rm cp}-\tau_{\rm cz}=25\pm4$\,My is
almost insensitive to the required solar age,
as seen in Table~\ref{tab:1} from the comparison
of the P-model Sp ($resp.$ Z-model Sz) of age 4530\,My
with the P-model SP ($resp.$ Z-model SZ) of age 4660\,My.
At solar age $t_\odot=4530$\,My ($resp.$ $t_\odot=4660$\,My)
the differences between the models Sp and Sz ($resp.$ SP and SZ) are
below $10^{-4}$ i.e. the level of accuracy of the calibration.
All these results validate our analysis.
\begin{figure}
\centerline{
\psfig{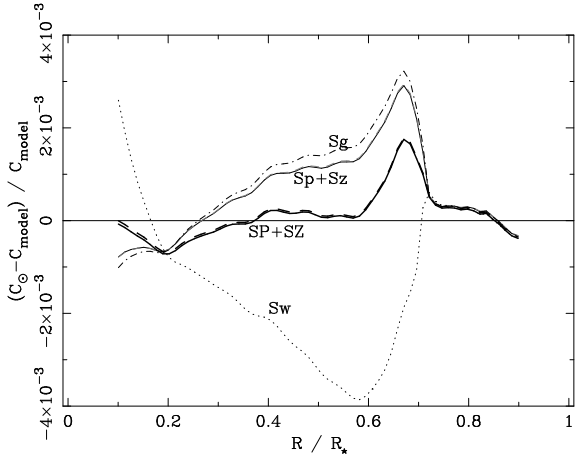}
}
\caption{
Relative difference in sound speed between the Sun and models,
Sp (thin, full), Sz (thin, dashed), SP (heavy, full),
SZ (heavy, dashed), Sg (dash-dot) and Sw (dotted).
}\label{fig:vson}
\end{figure}

As a matter of comparisons, we have calibrated the solar Z-models
Sg and Sw for ages
$t_\odot=4500$\,My (Guenther \& Demarque \cite{gd97}) and 
$t_\odot=5100$\,My (Weiss \& Schlattl \cite{ws98}), respectively. 
Figure~\ref{fig:vson} shows the relative difference between the sound
speed of models and the seismic
sound speed experimental results of Turck-Chi\`eze et al.
(\cite{t97}). The best fit appears to be for
the models SP (and SZ), of age $t_\odot=t_{\odot\rm h}=4660$\,My, i.e. the
helioseismic solar age of Dziembowski et al. (\cite{dfrs99}). This means that the physics used in solar models
leads to deficits  which are equivalent to an evolution slower than  indicated
by  $t_{\odot\rm m}$.

\section{Discussion and conclusions}\label{sec:dis}
Fixing the physics we have compared calibrated solar models initialized with a
 homogeneous zero-age main sequence model (Z-model) and with a
 pre-main sequence model (P-model). We have verified that the
 models which evolved from these different initial models merge
 into the same structure
 about 10\,My after the end of the convective core phase, if the 
 times, $t_{\rm cal}$, of evolution
 used in each calculation, are shifted by an amount of 25\,My.
That similarity allows us to connect $t_{\rm cal}$ to the solar age, i.e.
the time it has taken the Sun to evolve from the time (ZAMS),
where nuclear reactions
{\em just begin} to dominate gravitation as the primary energy source,
to present day. For the calibrated Z-model we found
$t_{\rm cal}=t_\odot$, while for a P-model
$t_{\rm cal}=t_\odot+25$\,My. We emphasized the fact that
these two relations between $t_{\rm cal}$ and $t_\odot$
are only valuable
when the zero-age reference corresponds to the time when
the nuclear reactions {\em just begin} to be
the primary energy source. If the epoch of the ZAMS is defined at the
instant where 99\%, instead of 50\%, of energy comes from nuclear reactions,
the shift becomes $50$\,My, then, for a P-model
$t_{\rm cal}=t_\odot+50$\,My and
$t_{\rm cal}=t_\odot+25$\,My for a Z-model.
Though $25$\,My represent only 0.5\% of the solar age, it is of the same
order of accuracy as present day solar models.
The basic idea of our analysis is based on the similarity between
the models soon after the convective core exhaustion;
it is useless for stars of masses greater than about $1.2\,M_\odot$,
which exhibit a convective core on the main sequence.
Another analysis is needed for these stars,
that may be of importance in asteroseismology to differentiate
between age and $t_{\rm cal}$ for modeling
the COROT targets (Baglin \cite{bc98}).

\begin{acknowledgements}
We thank the referee Dr. A. Weiss for valuable comments.
This work has been performed using the computing facilities 
provided by the OCA program
``Simulations Interactives et Visualisation en Astronomie et M\'ecanique 
(SIVAM)''.
\end{acknowledgements}

\end{document}